\documentclass[useAMS,usenatbib]{mn2e}
\usepackage{amssymb}
\usepackage{natbib}
\usepackage{times}
\usepackage{graphics,epsfig}
\usepackage{graphicx}
\usepackage{amsmath}
\usepackage{amssymb}
\usepackage{comment}
\usepackage{supertabular}

\usepackage{graphicx}
\usepackage{amsmath}
\usepackage{color}

\newcommand{\kms}{km~s$^{-1}$}
\newcommand{\arcs}{$^{\prime\prime}$}

\newcommand{\ha}{\ensuremath{{\rm H}\alpha}\ }

\newcommand{\mspcc}{\ensuremath{\rm M_{\odot} pc^{-3} }}
\newcommand{\HI}{H{\sc i}}
\newcommand{\HII}{H{\sc ii}}

\title[Thick disc molecular gas fraction in NGC 6946]{Thick disc molecular gas fraction in NGC 6946}
\author [N. N. Patra]{	Narendra Nath Patra$^{1}$ \thanks {E-mail: narendra@rri.res.in} \\
	$^{1}$ Raman Research Institute, C. V. Raman Avenue, Sadashivanagar, Bengaluru 560080, India\\
}
\date {}
\begin{document}
\maketitle

\begin{abstract}

Several recent studies reinforce the existence of a thick molecular disc in galaxies along with the dynamically cold thin disc. Assuming a two-component molecular disc, we model the disc of NGC 6946 as a four-component system consisting of stars, \HI, thin disc molecular gas, and thick disc molecular gas in vertical hydrostatic equilibrium. Following, we set up the joint Poisson-Boltzmann equation of hydrostatic equilibrium and solve it numerically to obtain a three-dimensional density distribution of different baryonic components. Using the density solutions and the observed rotation curve, we further build a three-dimensional dynamical model of the molecular disc and consecutively produce simulated CO spectral cubes and spectral width profiles. We find that the simulated spectral width profiles distinguishably differ for different assumed thick disc molecular gas fractions. Several CO spectral width profiles are then produced for different assumed thick disc molecular gas fractions and compared with the observed one to obtain the best fit thick disc molecular gas fraction profile. We find that the thick disc molecular gas fraction in NGC 6946 largely remains constant across its molecular disc with a mean value of $0.70 \pm 0.09$. We also estimate the amount of extra-planar molecular gas in NGC 6946. We find $\sim 50\%$ of the total molecular gas is extra-planar at the central region, whereas this fraction reduces to $\sim$ 15\% at the edge of the molecular disc. With our method, for the first time, we estimate the thick disc molecular gas fraction as a function of radius in an external galaxy with sub-kpc resolution.

\end{abstract}


\section{Introduction}

The molecular gas in galaxies acts as the raw fuel for star formation, and hence, it can significantly influence the physical and chemical evolution of the interstellar medium (ISM). The neutral hydrogen (\HI), on the other hand, plays the role of a long-term reservoir for star formation. The stars form out of molecular clouds, which itself is produced from the Cold Neutral Medium (CNM) phase of the atomic ISM. In that sense, the \HI~and the molecular gas closely related to star formation in spiral galaxies such as the star formation and the total gas surface densities (\HI+molecular) in these galaxies are found to be tightly correlated to each other \citep{schmidt59,kennicutt98b,bigiel08,leroy09b}. However, in smaller dwarf galaxies, this relation could be complicated as there is no significant detection of molecular gas despite great efforts \citep[see, e.g.,][]{taylor98b,schruba12}. Even though the star formation in these galaxies is found to correlate with the \HI~and, the CNM surface densities \citep{bigiel10b,roychowdhury09,roychowdhury11,roychowdhury14,roychowdhury17,patra16}. Hence, it is essential to investigate the ISM conditions of the \HI/molecular gas, which can directly influence the star formation. For example, in a recent study, \citet{bacchini19} has revealed that the traditional Kennicutt-Schmidt law (KS law, relation between gas surface density and the star formation rate surface density (SFR), \citet{kennicutt98b}) shows much less scatter when the SFR is compared with the volume density of the gas (\HI+molecular) instead of surface densities. This indicates that the volume density of the gas is more closely related to SFR than the surface densities, and hence, the determination of a three-dimensional distribution of the same would be imperative to understand the connection between gas and star formation.

The molecular disc in galaxies is thought to be a dynamically cold component settled very close to the midplane, having a very low kinetic temperature (a few tens of Kelvin). This demands the scale height, and the physical extent of the molecular gas in galaxies would be restricted to much lower heights as compared to the \HI~disc. However, many direct observations of the edge-on galaxy NGC 891 reveal a substantial amount of extraplanar gas and dust reaching up to heights of several kiloparsecs from the midplane. For example, in a deep Westerbork Synthesis Radio Telescope (WSRT) observation, \citet{fraternali05} detected a significant amount of extraplanar \HI~with a scale height of $\sim$ 3 kpc \citep[see also][]{swaters97,fraternali06}. Similarly, several others performed deep \ha~observations and traced the extraplanar diffused ionized gas in NGC 891 to a height as large as $\sim 4-5$ kpc (typical scale height of $\sim$ 1 kpc) \citep{dettmar90,rand90,hoopes99,kamphuis07,boettcher16}. Further, deep UV (scattered by dust) observations confirmed the presence of dust at heights $\sim$ 2 kpc from the midplane \citep{,howk2000,rossa04,kamphuis07b,seon14}. 

\citet{garcia-burillo92} detected molecular gas in NGC 891 at $\sim 1-1.4$ kpc above the midplane. The scale height of the molecular gas in a thin disc is expected to be a few hundred parsecs. Not only that, recent high spatial and spectral studies also provide adequate evidence that the molecular gas in galaxies can also exist in a thick disc with much larger scale heights.  For example, \citet{calduprimo13} used the data of 12 nearby large spiral galaxies from The HERA CO Line Extragalactic survey (HERACLES, \citet{leroy09a}) and The \HI~Nearby Galaxy Survey (THINGS, \citet{walter08}) to stack the \HI~and the CO spectra in these galaxies and found that the ratio, $\sigma_{HI}/\sigma_{CO} = 1.0 \pm 0.2$ with $\sigma_{HI}=11.9 \pm 3.1$ \kms. Later, \citet{mogotsi16} used the same data to identify the high SNR regions in both the \HI~and CO map and compared the velocity dispersions. They found a $\sigma_{HI}/\sigma_{CO} = 1.4 \pm 0.2$ with $\sigma_{HI} = 11.7 \pm 2..3$ \kms. These studies conclude that the molecular gas in galaxies exists in two phases/discs. One in a thin disc close to the midplane and has a low velocity dispersion (thin disc molecular gas). The other one exists in a more diffuse thick disc with a velocity dispersion the same as  $\sigma_{HI}$ (thick disc molecular gas). Due to the diffuse nature of the thick disc molecular gas, it is not detected in the individual spectra, whereas it shows its signature in a high SNR stacked spectrum.

The existence of this diffuse molecular disc in galaxies raises several vital questions. For example, what is the origin of this low density diffuse molecular gas? As they have the same velocity dispersion as the \HI, the thick disc molecular gas will have a similar scale height as the \HI~disc. In that sense it is unclear how this phase can survive at these heights from the midplane, where the metagalactic radiation is high. However, several earlier studies revealed the existence of dust in galaxies at a considerable height from the midplane \citep{howk2000,rossa04,kamphuis07b,seon14,shinn18,jo18}. This dust could, in principle, provide enough shielding to this diffuse molecular disc.  Nevertheless, the origin and physical conditions for such molecular phase sustenance at a considerable height are not yet clearly understood.

Several studies point towards a dynamical origin of the diffuse molecular gas in spiral galaxies. For example, it is proposed that due to the passage of the spiral density wave, there could be a spontaneous conversion of the atomic phase into the diffuse molecular phase \citep{blitz80,cohen80,scoville79,heyer15}. On the other hand, some studies suggest that the diffuse molecular gas already exists in the form of small molecular clouds, which are not detected within individual beams. The observed Giant Molecular Clouds (GMCs) in the arm then can be produced by dynamically induced coagulation due to the passage of the density wave \citep{scoville79,vogel88,hartmann01,kawamura09,koda11,miura12,pety13,koda16}. However, these studies cannot provide a satisfactory explanation of the substantial amount of extraplanar molecular gas at considerable heights. 

On the other hand, a number of high-resolution, sensitive observations revealed significant molecular outflows in galaxies due to enhanced star formation or Active Galactic Nuclei (AGN) activities. These molecular outflows can carry a significant amount of molecular gas and energy to produce the observed extraplanar molecular gas comfortably. For example, using CO observations, \citet{weiss99} detected a molecular superbubble in M82 having a diameter of $\sim 130$ pc, which has broken out into the Cicum Galactic Medium (CGM). Employing simple theoretical models, they infer that a supernovae rate of $\sim 0.001$ per year would be sufficient to produce this kind of superbubbles. \citet{walter02} observed molecular streamers in M82 using high resolution ($\sim 70$ pc) CO observations. These streamers are believed to be closely related to the starburst activity in the galaxy and can expel molecular gas to heights $\sim 1.2$ kpc. \citet{feruglio10} observed Mrk 231, a nearby quasar using the Plateau de Bure interferometer (PdBI), and found a substantial amount of molecular outflow from the central region. Likewise, several other studies also found evidence of starburst or AGN driven outflows leading the molecular gas to large heights \citep{sakamoto06,alatalo11,salak13,bolatto13,krieger19,salak20}. These mechanisms could be viable pathways for producing and maintaining the observed diffuse (extraplanar) molecular gas in galaxies.  

Though the detection of the diffuse molecular discs in galaxies has become less ambiguous with recent sensitive observations, its physical properties are largely unknown. For example, what is the fractional abundance of this component in galaxies? How does this fraction vary with radius? \citet{pety13} observed the nearby large spiral galaxy M51 with the Plateau de Bure Interferometer (PdBI) to map its molecular disc with an unprecedented spatial resolution of $\sim 40$ pc. They also used the data from the IRAM 30 m telescope to map the galaxy with a single-dish. They find that the single-dish observation recovers almost twice the flux detected by the PdBI interferometer. They concluded that the diffuse thick disc molecular gas is resolved out in the high spatial resolution interferometric observation, which is detected in low-resolution single-dish measurement. Comparing the fluxes detected in single-dish and the interferometric observations, they infer at least 50\% of the molecular gas in M51 is in the thick disc. However, this kind of measurement is global; i.e., it does not provide any information on how this fraction varies within a galaxy as a function of different physical conditions. Not only that, this kind of estimation requires high spatial resolution interferometric observation along with a single-dish measurement, which very often is not available simultaneously.

In this paper, we develop a method to estimate the thick disc molecular gas fraction as a function of radius in external galaxies and apply it to a nearby large spiral galaxy NGC 6946. We model the baryonic disc of the galaxy as a four-component system consisting of stars, \HI~and two molecular discs (thin and thick). We assume that these discs are in vertical hydrostatic equilibrium under their mutual gravity in the external force field of the dark matter halo. Under these assumptions, we set up the joint Poisson-Boltzmann equation of hydrostatic equilibrium and solve it numerically. With these solutions, we build a three-dimensional dynamical model of the molecular disc in NGC 6946 and produce CO spectral cubes and spectral width profiles equivalent to the observation. We find that these obtained spectral widths are sensitive to the assumed thick disc molecular gas fraction at every radius. Utilizing this characteristic, we produce a number of spectral cubes and hence spectral width profiles with different thick disc molecular gas fractions and compare it with the observed profile to constrain the thick disc molecular gas fraction in NGC 6946 as a function of radius.

\section{Modeling the disc of NGC 6946}

\begin{figure*}
\begin{center}
\begin{tabular}{c}
\resizebox{1.\textwidth}{!}{\includegraphics{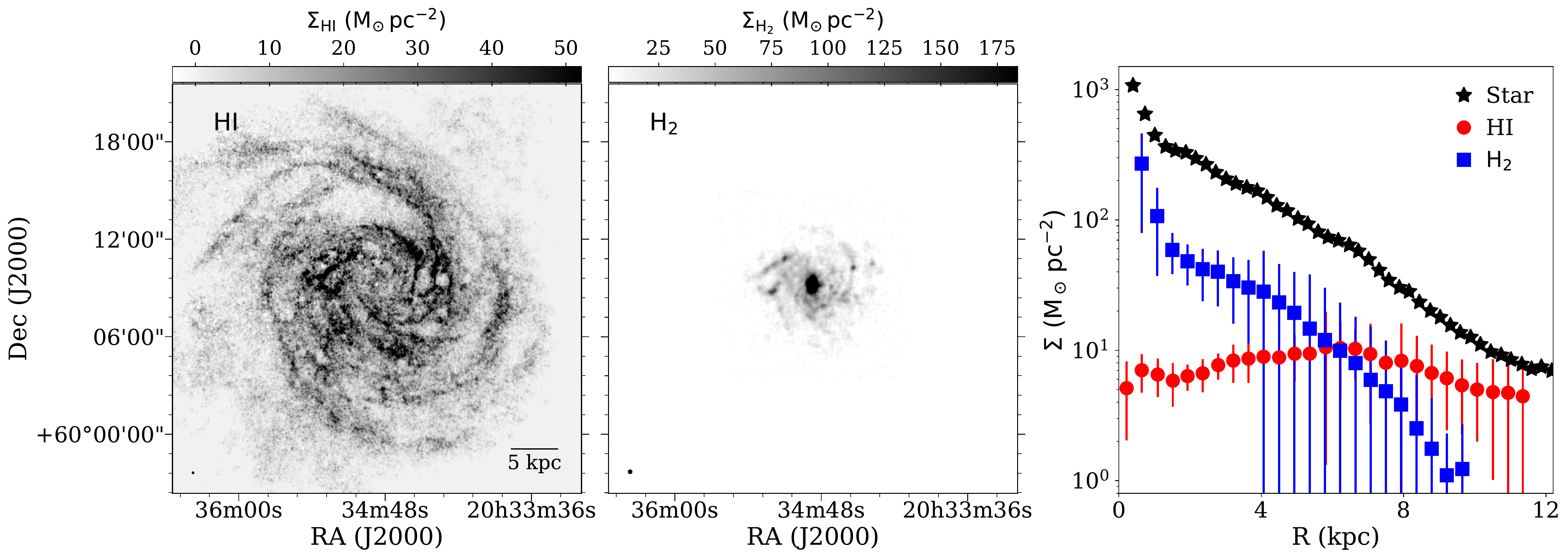}}
\end{tabular}
\end{center}
\caption{Left panel: The observed \HI~column density map of NGC 6946 \citep[THINGS survey,][]{walter08}. Middle panel: The observed column density map of the molecular gas \citep[HERACLES survey,][]{leroy09a}. Right panel: The surface density profiles for NGC 6946 \citep{leroy08,schruba11}. The black filled stars represent the stellar surface density profile, the solid red circles with error bars represent the \HI~surface density profile, whereas, the solid blue squares with error bars represent the molecular surface density profile. The observing beams are shown in the bottom left corners of the left and the middle panels. The color bars on the top of the left and the middle panels represent the column density in the units of $\rm M_{\odot} \thinspace pc^{-2}$. The left and the middles panel represents the same angular extent in the sky. As can be seen from these two panels, the \HI~disc extends to much larger radii as compared to the molecular disc.}
\label{sden}
\end{figure*}

As mentioned above, we construct several model molecular discs for NGC 6946 with different thick disc molecular gas fraction, $\rm f_{tk}$, and produce expected spectral width profiles for all of them. To do that, we model the baryonic disc in NGC 6946 as a four-component system consisting of stars, \HI, thick disc molecular gas, and thin disc molecular gas. We assume that all these discs are in vertical hydrostatic equilibrium. For simplicity, we consider all the discs are coplanar and concentric. With these assumptions, the Poisson's equation of hydrostatic equilibrium in cylindrical polar coordinate can be written as 

\begin{equation}
\frac{1}{R} \frac{\partial }{\partial R} \left( R \frac{\partial \Phi_{total}}{\partial R} \right) + \frac{\partial^2 \Phi_{total}}{\partial z^2} = 4 \pi G \left( \sum_{i=1}^{4} \rho_{i} + \rho_{h} \right)
\label{eq1}
\end{equation}

\noindent where $\Phi_{total}$ is the total potential due to all the baryonic components and the dark matter halo. $\rho_i$ represents the volume density of the disc components where {\it i} runs for stars ($\rho_s$), \HI~($\rho_{HI}$), thin disc molecular gas ($\rho_{H2,tn}$) and thick disc molecular gas ($\rho_{H2,tk}$). $\rho_h$ denotes the density of the dark matter halo. For this modeling, we consider the dark matter halo to provide a fixed potential (not live), which is determined by the mass-modeling of the observed rotation curve. 

The total potential, $\Phi_{total}$, is not a directly measurable quantity, and hence, Eq.~\ref{eq1} can not be solved in its present form. Instead, we use the Boltzmann equation, to replace the $\Phi_{total}$ with a more directly measurable quantity, vertical velocity dispersion. In hydrostatic equilibrium, the gradient in pressure will be balanced by the gradient in the total potential. Hence,

\begin{equation}
\frac{\partial }{\partial z} \left(\rho_i {\langle {\sigma}_z^2 \rangle}_i \right) + \rho_i \frac{\partial \Phi_{total}}{\partial z} = 0
\label{eq2}
\end{equation}

\noindent Where $(\sigma_z)_i$ is the vertical velocity dispersion of the $i^{th}$ disc component. Eq.~\ref{eq2} can be used to simplify Eq.~\ref{eq1} such as,

\begin{equation}
\begin{split}
{\langle {\sigma}_z^2 \rangle}_i \frac{\partial}{\partial z} \left( \frac{1}{\rho_i} \frac{\partial \rho_i}{\partial z} \right) &= \\ 
&-4 \pi G \left( \rho_s + \rho_{HI} + \rho_{H2,tn} + \rho_{H2,tk} + \rho_h \right)\\ 
&+ \frac{1}{R} \frac{\partial}{\partial R} \left( R \frac{\partial \Phi_{total}}{\partial R} \right)
\end{split}
\label{eq3}
\end{equation}

Assuming the \HI~rotation curve traces the total gravitational potential reasonably accurately, the last term in the above equation can be computed as 

\begin{equation}
{\left( R \frac{\partial \Phi_{total}}{\partial R} \right)}_{R,z} = {(v_{rot}^2)}_{R,z}
\label{eq4}
\end{equation}

Putting Eq.~\ref{eq4} in Eq.~\ref{eq3}, we get the final hydrostatic equilibrium equation as

\begin{equation}
\begin{split}
{\langle {\sigma}_z^2 \rangle}_i \frac{\partial}{\partial z} \left( \frac{1}{\rho_i} \frac{\partial \rho_i}{\partial z} \right) &= \\
&-4 \pi G \left( \rho_s + \rho_{HI} + \rho_{H2,tn} + \rho_{H2,tk} + \rho_h \right)\\ 
&+ \frac{1}{R} \frac{\partial}{\partial R} \left( v_{rot}^2 \right)
\end{split}
\label{eq5}
\end{equation}

Eq.~\ref{eq5} represents four coupled second-order partial differential equations in $\rho_{s}$, $\rho_{HI}$, $\rho_{H2,tn}$, and $\rho_{H2,tk}$. The solutions of this equation at a radius will provide the density distribution of the different disc components as a function of the vertical height ($z$) from the midplane. To obtain a complete three-dimensional density distribution of the disc components, one needs to solve Eq.~\ref{eq5} at all radii. However, several input parameters are needed to solve Eq.~\ref{eq5}, as described below.

\subsection{Input parameters}

\subsubsection{Gravity terms}
In vertical hydrostatic equilibrium, the gravity will be balanced by the pressure. In that sense, the baryonic surface densities are one of the primary inputs which provide a significant amount of gravity into Eq.~\ref{eq5}. \citet{leroy09a} observed NGC 6946 using the 30-m IRAM telescope as part of the HERACLES survey with a spatial and spectral resolution of 13.4\arcs~and 2.6 \kms~respectively. 13.4\arcs~at the distance of NGC 6946 (5.9 Mpc, \citet{karachentsev04}) translates into a linear scale of $\sim 400$ pc. \citet{schruba11} constructed the molecular surface density profile of NGC 6946 by adopting a spectral stacking technique. All the line-of-sight CO spectra within rings of width 15\arcs~(as obtained by the tilted ring fitting of the \HI~data, see \citet{deblok08} for more details) are stacked after shifting their centers to a common velocity. These resulting stacked spectra for different radial bins have much higher S/N than any individual line-of-sight spectrum. The molecular surface density profile is then estimated by fitting these stacked spectra. The surface density profile of the atomic gas is obtained by averaging the \HI~intensity distribution (as obtained by the THINGS survey) within the same tilted rings \citep[see][for more details]{schruba11}. A correction factor of 1.4 is applied to the surface density profile of the atomic gas to account for the presence of the cosmological Helium in the ISM. We adopt the stellar surface density profile of NGC 6946 as estimated by \citet{leroy08} using 3.6$\mu m$ data from the Spitzer Infrared Nearby Galaxy Survey \citep[SINGS,][]{kennicutt03}. In Fig.~\ref{sden}, we show the \HI~(left panel) and the molecular (middle panel) maps of NGC 6946 from the THINGS and the HERACLES survey, respectively. The angular extent in both the panels is the same. Hence, it can be seen from the figure, the \HI~disc in NGC 6946 extends to much larger radii as compared to the molecular disc. The respective surface density profiles are shown in the rightmost panel.

Another significant source of gravity is the dark matter halo. The dark matter halo might provide a considerable amount of gravity in the vertical direction within the baryonic disc. Specifically, it could be relevant at outer radii, where the gas discs flare significantly. For NGC 6946, we use the mass-model as derived by \citet{deblok08}. \citet{deblok08} fit the high-resolution rotation curve as obtained from the THINGS data with both an NFW and a pseudo-isothermal (ISO) halo. However, for NGC 6946, an ISO halo found to represent the observed rotation curve better than an NFW one. They found a characteristic core density, $\rho_0 = 45.7 \times 10^{-3}$ \mspcc~and a core radius, $R_c = 3.62$ kpc (see their Table 3 and 4 for more details). We use these two parameters to describe the dark matter halo density distribution in NGC 6946, which can be given as 

\begin{equation}
\label{eq_iso}
\rho_h(R) = \frac {\rho_0}{1 + \left(\frac{R}{R_s}\right)^2}
\end{equation}

\noindent where, $\rho_h(R)$ represents the dark matter halo density as a function of the radius. 

\subsubsection{Rotation curve}
\begin{figure}
\begin{center}
\begin{tabular}{c}
\resizebox{0.4\textwidth}{!}{\includegraphics{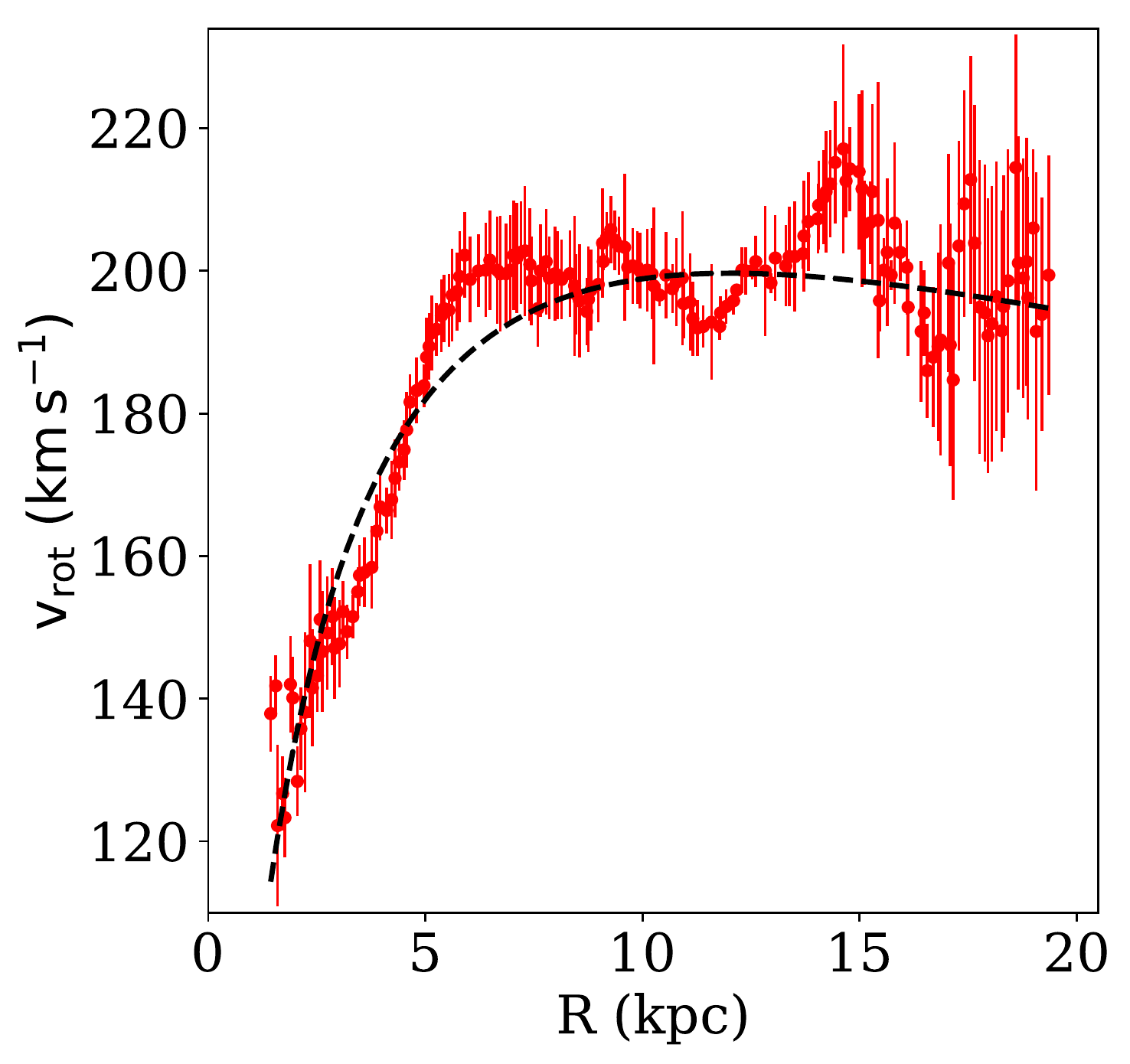}}
\end{tabular}
\end{center}
\caption{The rotation curve of NGC 6946. The solid red circles with error bars represent the observed rotation curve as derived by a tilted ring model fitting to the high resolution \HI~data from the THINGS survey \citep{deblok08}. The black dashed line represents a Brandt-profile fit to the data.}
\label{rotcur}
\end{figure}

Next, the observed rotation curve acts as another essential input parameter, which is necessary to calculate the radial term, i.e., the last term on the RHS of Eq.~\ref{eq5}. \citet{deblok08} used the high-resolution \HI~data from the THINGS survey to perform a tilted ring model fitting to the two-dimensional velocity field as obtained by a Gaussian-Hermite method. In Fig.~\ref{rotcur}, we show the rotation curve for NGC 6946. As can be seen from the figure, the rotation velocity increases as a function of the radius at the central region and flattens at $R \gtrsim 7$ kpc. It can also be seen from the figure that due to very high spatial resolution, the rotation curve exhibits significant fluctuations. However, in Eq.~\ref{eq5}, one needs to calculate the first derivative of the squared rotation velocity to calculate the radial term. A sudden variation/jump in the rotation curve might lead to an unphysical value of this derivative and diverge the solutions of the hydrostatic equation. To avoid the same, we fit the rotation curve with a commonly used Brandt profile \citep{brandt60},

\begin{equation}
v_{rot} (R) = \frac{V_{max}\left(R/R_{max} \right)}{\left(1/3 + 2/3 \left(\frac{R}{R_{max}}\right)^n\right)^{3/2n}}
\label{brandt}
\end{equation}

\noindent where $v_{rot}$ represents the observed rotation velocity at any radius, $V_{max}$ is the maximum attained velocity, $R_{max}$ is the radius at which $V_{max}$ is achieved. $n$ represents an index which signify how fast or slow the rotation curve rises as a function of the radius. For NGC 6946, a Brandt profile fit (black dashed curve in the figure) results in, $V_{max} = 199.7 \pm 0.5$ \kms, $R_{max} = 12.0 \pm 0.6$ kpc, and, $n=0.69 \pm 0.05$. As can be seen from the figure, the Brandt profile fit in the central region is not steep enough. As a result, it appears to overestimates the value of $R_{max}$. However, we note that the radial term does not contribute to Eq.~\ref{eq5} considerably. Such as a value of $R_{max} = 7$ kpc, or an exponential fit to the rotation curve \citep{boissier03c,leroy08} alters the solutions by less than one percent. Hence, we use the Barndt profile fit parameters to parametrize the rotation curve and solve the hydrostatic equation.

\subsubsection{Velocity dispersion}

The velocity dispersion of different baryonic components is another critical input parameters required to solve Eq.~\ref{eq5}. The gravity on any elemental volume in the disc is contributed by all the four disc components and the dark matter halo. In that sense, a change in the surface density of any component is adjusted by all the baryonic discs. On the other hand, the velocity dispersion of an individual disc component solely decides the pressure. Thus the velocity dispersion can significantly influence the vertical structure of the baryonic discs and should be estimated precisely.

Direct measurement of the stellar velocity dispersion is difficult even with modern-day telescopes. As a result, the stellar velocity dispersion for NGC 6946 is not available. Instead, we adopt an analytical approach to estimate the stellar velocity dispersion in NGC 6946, assuming its stellar disc to be a single-component system in vertical hydrostatic equilibrium. We use the corresponding analytical expression from \citet{leroy08} (see their Appendix B) to calculate the stellar velocity dispersion. It should be mentioned here that this calculation does not include the gravity due to gas discs and hence underestimates the stellar velocity dispersion. Nonetheless, the velocity dispersion of the stars does not influence the density distribution of the gas discs considerably \citep[see, e.g.,][]{banerjee11}. As we are primarily interested in the distribution of the molecular gas in NGC 6946, an analytical approximation to the stellar velocity dispersion is adequate for solving the hydrostatic equation in this case.

However, unlike stellar velocity dispersion, the gas velocity dispersion in galaxies could be estimated using spectral line observations. The early low-resolution \HI~observations in spiral galaxies revealed an \HI~velocity dispersion ($\sigma_{HI}$) of 6-13 \kms~\citep{shostak84,vanderkruit84,kamphuis93}. \citet{tamburro09} extensively studied the second moment of the high-resolution \HI~spectral cubes (MOM2) in a large number of spiral galaxies from the THINGS survey. They found an average $\sigma_{HI}$ of $\sim$ 10 \kms~at the optical radius ($r_{25}$) of the galaxies. Later, \citet{ianjamasimanana12} used the same sample to stack the line-of-sight \HI~spectra to generate high SNR {\it super-profiles}. They found,  $\sigma_{HI} = 12.5 \pm 3.5$ \kms~($\sigma_{HI} = 10.9 \pm 2.1$ \kms~for galaxies with inclination less than 60$^o$). However, though, all these studies found a significant variation in $\sigma_{HI}$ within and across the galaxies. 

\begin{figure}
\begin{center}
\begin{tabular}{c}
\resizebox{0.48\textwidth}{!}{\includegraphics{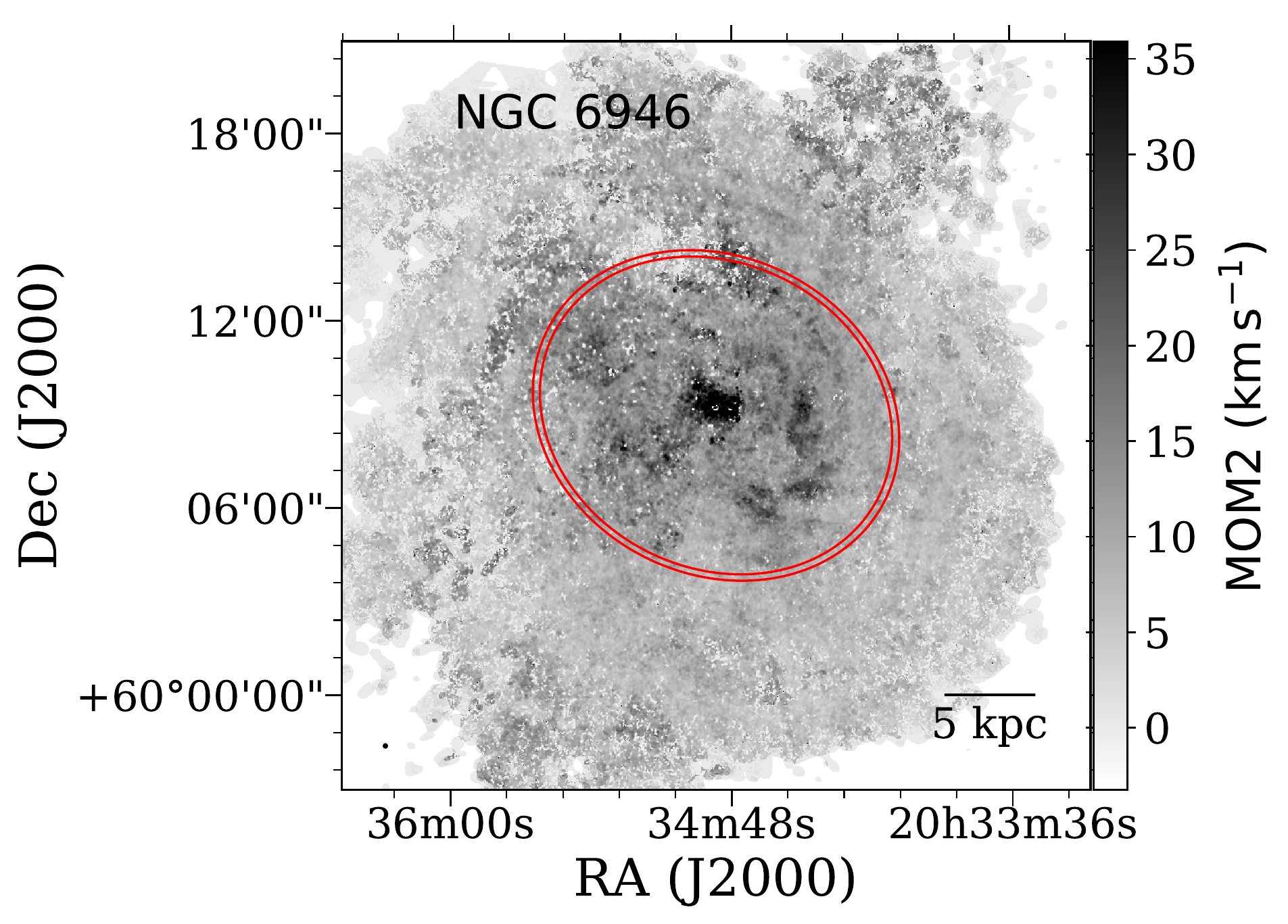}}\\
\resizebox{0.4\textwidth}{!}{\includegraphics{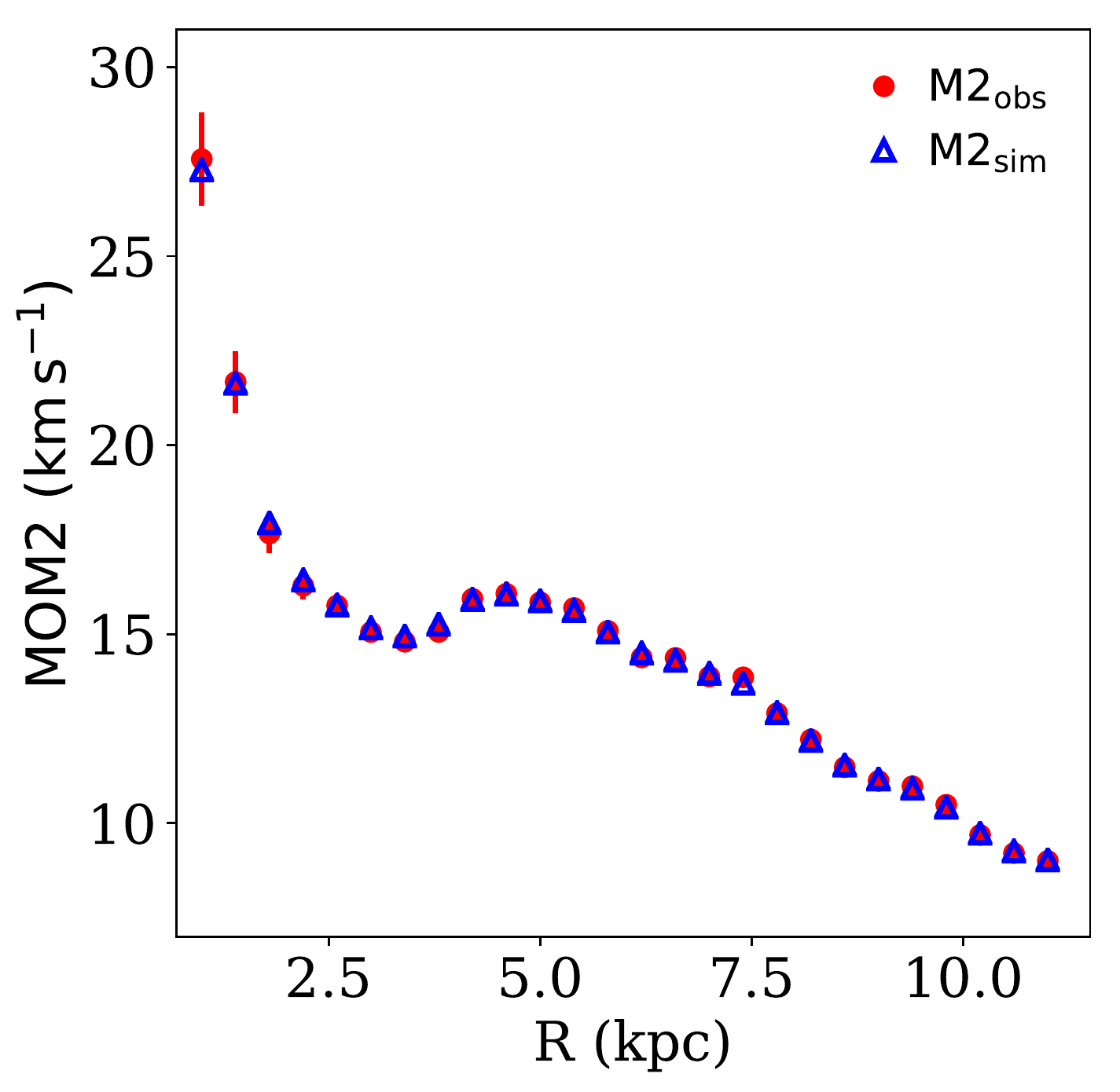}}
\end{tabular}
\end{center}
\caption{Top panel: Shows the MOM2 map of NGC 6946 as obtained using the THINGS survey data. The grayscale represents the observed MOM2 calculated by taking the second moment of the \HI~spectral cube. The red elliptical annular region represents a typical ring at a galactocentric distance of $\sim 10$ kpc, within which the observed MOM2 values are averaged to estimate the MOM2 profile. The telescope beam is shown by the solid black circle at the left bottom corner of the panel. The bottom panel shows the MOM2 profile for NGC 6946 using the map at the top panel. The solid red circles with error bars represent the observed MOM2, whereas the empty blue triangles are the simulated MOM2 with observed MOM2 as input $\sigma_{HI}$.}
\label{sig_mo2}
\end{figure}

To incorporate this variation of $\sigma_{HI}$ within NGC 6946, we use its MOM2 map as obtained from the THINGS survey to estimate the $\sigma_{HI}$ profile. In Fig.~\ref{sig_mo2} top panel, we show the corresponding MOM2 map. The MOM2 profile is determined by averaging MOM2 values within radial annuli of widths $\sim$ twice the beam size. A typical annulus at $R=10$ kpc is shown in the figure (red ellipses). Thus produced MOM2 profile is shown in the bottom panel of the figure (solid red circles with error bars). The error bars indicate one-sigma scatter in MOM2 within an annulus normalized by the square root of the number of independent beams within that annulus.

As the MOM2 is the intensity weighted $\sigma_{HI}$ along a line-of-sight, it always overestimates the intrinsic \HI~velocity dispersion (due to spectral blending). To investigate the degree of spectral blending in NGC 6946, we produce simulated MOM2 using the observed MOM2 as input $\sigma_{HI}$ in Eq.~\ref{eq5}. Using the density solutions of Eq.~\ref{eq5} (see \S 3 and \S 4) and the observed rotation curve, we build a three-dimensional dynamical model of the \HI~disc in NGC 6946. This \HI~disc is then inclined to the observed inclination, projected to the sky plane, and convolved with the telescope beam to produce an \HI~spectral cube. A simulated MOM2 profile is then computed using this spectral cube, which is shown by the empty blue triangles in the bottom panel of Fig.~\ref{sig_mo2}. As can be seen, the simulated and the observed MOM2 matches with each other very well, indicating a minimal spectral blending \citep[see][for more details]{patra20c}. Hence, for NGC 6946, we use the observed MOM2 profile as the intrinsic $\sigma_{HI}$ to solve Eq.~\ref{eq5}. 

In the Milky Way, observationally, it has been found that the low-mass molecular clouds show higher velocity dispersion ($\sigma_{CO} \sim 9$ \kms) as compared to the high-mass clouds ($\sigma_{CO} \sim 6.6$ \kms) \citep{stark84,stark05}. \citet{calduprimo13} stacked line-of-sight CO and \HI~spectra within radial bins in 12 nearby large spiral galaxies using the data from the HERACLES and the THINGS survey respectively. They estimate the spectral width profiles in these galaxies by fitting these stacked spectra by single-Gaussian functions. Thus they find a $\sigma_{HI}/\sigma_{CO} = 1.0 \pm 0.2$. Later, \citet{mogotsi16} investigated individual high SNR CO spectra in the same sample of galaxies and found a $\sigma_{HI}/\sigma_{CO} = 1.4 \pm 0.2$ with a $\sigma_{CO} = 7.3 \pm 1.7$ \kms. These studies conclude that the molecular disc has two components, a thin disc that is bright and detected in the individual spectra of \citet{mogotsi16} and has a velocity dispersion of $\sim 7$ \kms. The other component is diffuse and has a much lower density, which is not detected in the individual spectrum of \citet{mogotsi16} but discovered in the stacked spectra of \citet{calduprimo13}. This component has a velocity dispersion roughly equivalent to that of the \HI. Given these results, we use a $\sigma_{CO} = 7$ \kms~for the thin disc molecular gas and a $\sigma_{CO} = \sigma_{HI}$ for the thick disc molecular gas to solve Eq.~\ref{eq5}.

\subsubsection{Thick disc molecular gas fraction}

The fractional amount of molecular gas in the thin or the thick disc of NGC 6946 is not known, which we try to estimate in this work. In fact, a handful of studies to date directly estimates the thin or thick disc molecular gas fraction in galaxies. \citet{pety13} used the observations from the PdBI interferometer and the 30 m single-dish IRAM telescope to compare the detected CO fluxes in the nearby spiral galaxy M51. They found a flux discrepancy (higher flux recovery in the single-dish measurement) in both the observations and attributed it to the diffuse nature of the thick disc molecular gas, which has resolved out in the interferometric observation. They concluded that at least 50\% of the molecular gas in M51 is in the thick disc. However, this kind of study requires both the interferometric and single-dish observations. Not only that, but the interferometric observation also must have a high spatial resolution (was $\sim 40$ pc for M51). Such observational requirements are expensive and very often not possible to achieve except for very nearby galaxies. Here we use a different approach to estimate the thick/thin disc molecular gas fraction in NGC 6946, which could be applied more universally to other galaxies using existing observations. For this study, we do not consider the thick disc molecular gas fraction ($\rm f_{tk}$) in NGC 6946 as constant. Instead, we solve the hydrostatic equilibrium equation allowing the $\rm f_{tk}$ to vary from 0.05 to 0.95 in steps of 0.05. For all $\rm f_{tk}$ values, we solve Eq.~\ref{eq5} and generate the density solutions. Using these solutions, we produce observables for all the cases and compare them individually with the observed data to estimate the $\rm f_{tk}$ profile, which explains the observation best.

\section{Solving the hydrostatic equation}

Eq.~\ref{eq5} represents four coupled second-order partial differential equations in $\rho_s$, $\rho_{HI}, \rho_{H2,tn}$, and $\rho_{H2,tk}$. This equation cannot be solved analytically even for a two-component system. Hence, we solve Eq.~\ref{eq5} numerically using 8$^{th}$ order Runge-Kutta method as implemented in the python package {\tt scipy}. We adopt a similar strategy as used in our previous works \citep{banerjee11b,patra18a,patra19b}. We refer the readers to these papers for a detailed description of the numerical method we use to solve the hydrostatic equilibrium equation. Here we describe the basic strategy in brief.

As Eq.~\ref{eq5} is a second-order differential equation, one needs at least two initial conditions to solve the equation. We choose the initial conditions as follows.

\begin{equation}
\left( \rho_i \right)_{z = 0} = \rho_{i,0} \ \ \ \ {\rm and} \ \ \ \left(\frac{d \rho_i}{dz}\right)_{z=0} = 0
\label{init_cond}
\end{equation}

Due to the symmetry of the problem, there should be a density maxima (and hence maximum gravitational potential) at the midplane. This results in the second condition in the above equation. However, the first condition requires prior knowledge of the midplane density ($(\rho_i)_{z=0}$) for all the disc components, which is not available. We solve this problem by using the knowledge of the observed surface density. While solving Eq.~\ref{eq5} for any component, we start with a trial midplane density, $\rho_{t,0}$ and generate the solutions, $\rho_t(z)$. This solution is then integrated to obtain the trial surface density, $\Sigma_t = 2 \times \int \rho_t(z) dz$. This $\Sigma_t$ is then compared with the observed surface density to update the trial $\rho_{t,0}$ in the next iteration. Thus, we iteratively approach to a right $\rho_{t,0}$, which produces the observed surface density within 0.1\% accuracy.

Eq.~\ref{eq5} represents four coupled differential equations, which should be solved simultaneously. However, as the exact mathematical form of the gravitational coupling is not known, we adopt an iterative approach to introduce the coupling while solving the equations. In the first iteration, we solve individual equations (for stars, \HI, thin disc molecular gas, and thick disc molecular gas) without any coupling. This results in density solutions that did not consider the gravity of the other disc components. In the successive iterations, while solving for an individual component, we introduce the density solutions of the other disc components from the previous iteration into the gravity term. Thus, the density solutions grow slightly better than what was obtained in the previous iteration. We continue this iterative method until the solutions of all the components converge to better than 0.1\% accuracy. We emphasize that all the codes are implemented using MPI based parallel coding for faster computation. For more details, we refer the readers to \citet{narayan02b,banerjee11,patra18a,patra19b,patra20a,patra20c,patra20b}.

\section{Results and discussion}

\subsection{Density solutions}

Adopting the technique mentioned above, we solve Eq.~\ref{eq5} for NGC 6946 in every 100 pc at radius, $0.7 \geq R \geq 9.5$ kpc. As the spatial resolution of the molecular data is $\sim 400$ pc, a radial sampling of 100 pc should be adequate to capture the density variation of the molecular disc in full details. Due to the enhanced star formation and other energetic activities at the central regions of galaxies, a hydrostatic equilibrium condition might not be prevalent in these regions. As a signature, the spectral width in these regions can attain very high values due to higher turbulent and non-circular motions. For NGC 6946, the \HI~spectral width at the central region reaches up to $\sim 50$ \kms. To avoid this problem, in NGC 6946, we exclude a central region of $\sim 700$ pc and do not solve Eq.~\ref{eq5} at $R \leq 700$ pc.

\begin{figure}
\begin{center}
\begin{tabular}{c}
\resizebox{0.45\textwidth}{!}{\includegraphics{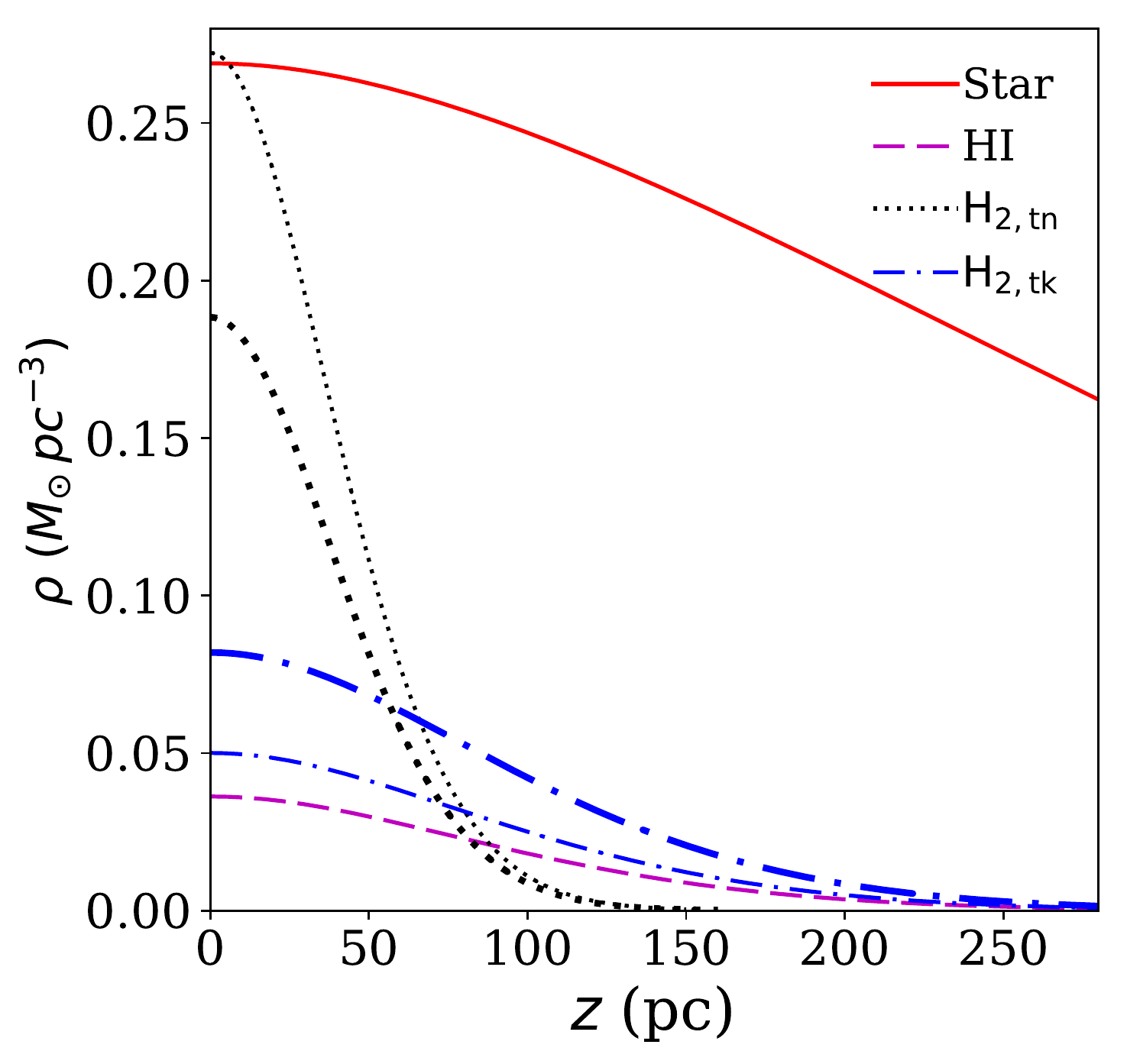}}
\end{tabular}
\end{center}
\caption{Sample density solutions of the hydrostatic equilibrium equation for NGC 6946 at R=3 kpc. The thin lines represent the solutions for a thick disc gas fraction of 0.3, whereas the thick lines represent solutions for a thick disc molecular gas fraction of 0.5. The solid red line represents the density solution for stars; the magenta dashed line depicts the solution for \HI. The dotted and the dashed-dotted lines denote the density solutions for the thin and the thick disc molecular gas. As can be seen from the figure, the thick disc molecular gas extends further in the vertical direction as compared to the thin disc.}
\label{densol}
\end{figure}

In Fig.~\ref{densol}, we plot sample solutions of Eq.~\ref{eq5} for NGC 6946 at a radius of 3 kpc. As can be seen from the figure, the stellar disc (solid red line) reaches much larger heights as compared to the \HI~(magenta dashed line) or the molecular discs (thin dotted and dashed-dotted lines). It can also be seen from the figure that the molecular gas in the thick disc (thin dashed-dotted lines) extends to much larger heights than the molecular gas in the thin disc (thick dotted lines). These density solutions (thin lines) are estimated for a $\rm f_{tk} = 0.3$. For comparison, we also plot the density solutions of the molecular discs for a $\rm f_{tk} = 0.5$ (thick lines). As can be seen, for a different $\rm f_{tk}$, the density distribution of the molecular discs in the vertical direction changes significantly. Consequently, it can produce very different observational signatures in the surface density maps or in the CO spectral cubes. We note that for an individual component in hydrostatic equilibrium without any coupling, the density solutions follow a $sech^2$ law \citep[see, e.g.,][]{bahcall84a,bahcall84b}. However, due to coupling between the disc components, the exact solutions deviate from a $sech^2$ law and behave more like a Gaussian function.

\subsection{Effects of lags}

\begin{figure}
\begin{center}
\begin{tabular}{c}
\resizebox{0.45\textwidth}{!}{\includegraphics{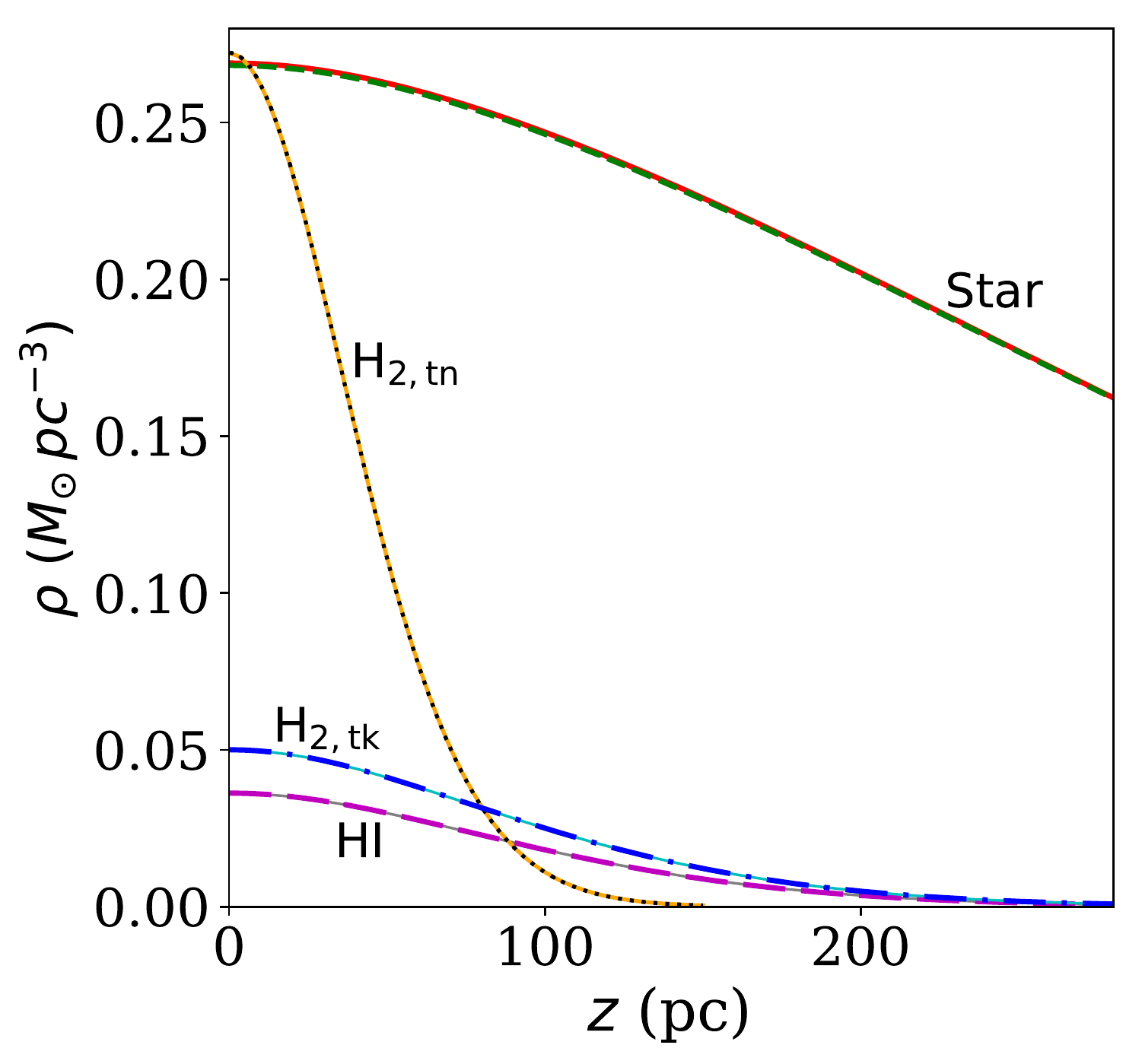}}
\end{tabular}
\end{center}
\caption{Density solutions at R=3 kpc for a lagging disc with a rotational velocity gradient of 30 $\rm km \thinspace s^{-1} \thinspace kpc^{-1}$. For comparison, the solutions for a disc without any lag is also plotted. The solid red, dashed magenta, dotted black, and dashed-dotted blue lines represent the density solutions for stars, \HI, thin, and thick disc molecular gas for a galactic disc with no lag. The dashed green, solid grey, solid orange, and solid cyan lines represent the same for a lagging disc. As can be seen, the solutions of Eq.~\ref{eq5} do not change considerably due to the lag in the rotation curve in the vertical direction.}
\label{densol_lag}
\end{figure}

While solving Eq.~\ref{eq5}, the rotation curve was assumed to be unchanged in the vertical direction. This forces the radial term (the last term on the RHS of Eq.~\ref{eq5}) to be constant as a function of $z$. However, several observational studies revealed the existence of a significant amount of gas at large altitudes from the midplane. This gas is often referred to as extra-planar gas in literature \citep{swaters97,fraternali02,fraternali05,zschaechner15,zschaechner15b,
marasco19,haffner09,bizyaev17,levy19}. The extra-planar gas was also found to exhibit distinct kinematic signature, as they rotate slower (lag) than the disc gas close to the midplane. In fact, this vertical lag in the rotation is commonly used to identify and isolate the extra-planar gas in galaxies \citep[see, e.g.,][]{swaters97,schaap00,chaves01,fraternali02,fraternali05,zschaechner15,
zschaechner15b,vargas17,marasco19}. In such cases, a barotropic steady-state hydrostatic model \citep[such as, by][]{barnabe06,fraternali06,marinacci10} of galactic discs would fail to capture the observed vertical lag in rotation. Instead, other mechanisms, such as anisotropic velocity dispersions \citep{marinacci10} or a rotating corona \citep{marinacci11}, would be required to account for the observed lags. In such realistic circumstances, the assumption of a constant radial term as a function of $z$ will not be valid fully. In fact, in NGC 6946, \citet{boomsma08} detected a considerable amount of extra-planar neutral gas, which exhibits the signature of a stong lag in the vertical direction. Hence, to investigate the effect of this lag on the vertical density distribution, we solve Eq.~\ref{eq5}, introducing appropriate variation in the radial term as a function of $z$. For NGC 6946, the magnitude of the lag is not well quantified. Nevertheless, several previous studies have estimated the vertical lag in a number of galaxies. For example, in a recent study, \citet{marasco19} used sensitive \HI~data from the Hydrogen Accretion in LOcal GAlaxieS (HALOGAS) survey to characterize the extra-planar \HI~in 15 nearby galaxies. They found a typical vertical gradient in rotational velocity of $\rm \sim - (10.0 \pm 3.7) \ km  s^{-1} kpc^{-1}$. However, several studies indicate that this lag could be as large as $\rm \sim - 30 \ km  s^{-1} kpc^{-1}$ for some galaxies \citep[see, e.g.,][]{zschaechner15,zschaechner15b}. Adopting a conservative approach, we use a lag of $\rm -30 \ km \  s^{-1} kpc^{-1}$ in Eq.~\ref{eq5} and examine its effect on the vertical density distribution. In Fig.~\ref{densol_lag} we show the resulting density solutions. As can be seen, the vertical gradient in the rotation velocity (or lag) does not have any meaningful effect ($\lesssim 0.3\%$) on the density solutions of Eq.~\ref{eq5}. This is because the radial term only marginally contributes to Eq.~\ref{eq5}. We chose to use solutions for a non-lagging disc in NGC 6946 for further analysis.

\subsection{Scale height measurements}

\begin{figure}
\begin{center}
\begin{tabular}{c}
\resizebox{0.45\textwidth}{!}{\includegraphics{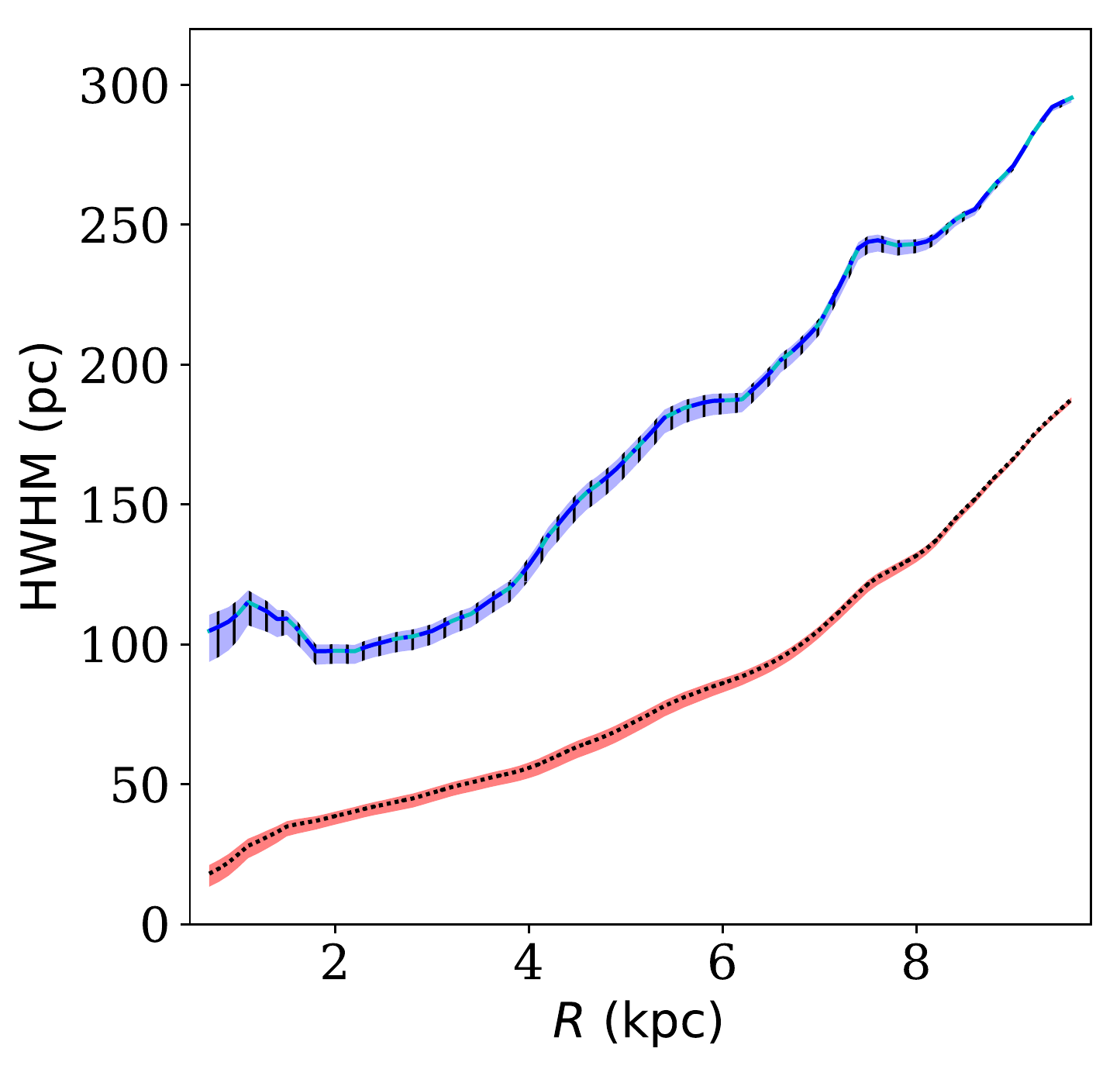}}
\end{tabular}
\end{center}
\caption{Show the scale height (HWHM) of different disc components as a function of radius. The red shaded region demarcates the scale height span of the thin disc molecular gas for $\rm f_{tk}$ between 0.3-0.9. The blue shaded region represents the scale heights of the thick disc molecular gas for the same range of $\rm f_{tk}$. The hatched region shows the range of \HI~scale heights, which exactly matches that of the thick disc molecular gas since they have the same assumed velocity dispersion. The black dotted line represents the thin disc scale height for an $\rm f_{tk} = 0.7$. The solid blue line and the long-dashed cyan line indicate the scale heights for the thick molecular and the \HI~discs, respectively (for $\rm f_{tk} = 0.7$).}
\label{sclh}
\end{figure}

We further use the density solutions to estimate the \HI~and molecular scale heights in NGC 6946. The scale height is defined as the Half Width at Half Maxima (HWHM) of the vertical density distribution. In Fig.~\ref{sclh}, we show the scale heights for the thin (red shaded region) and thick (blue shaded region) molecular discs as well as for \HI~(hatched region). As can be seen, the thin disc molecular scale height varies between $\sim$ 20 pc to $\sim$ 180 pc, whereas the same can vary between $\sim 100-300$ pc for the thick disc and \HI. Due to the same assumed velocity dispersion, the thick disc molecular gas and the \HI~produces the same scale heights as a function of radius. The scale heights of all the disc components flares as a function of radius and show a strong trend. 

We find that our estimated molecular scale heights in NGC 6946 are $\sim$ factor of two higher than what is observed in the Galaxy (in the same radial range). The molecular scale height in the Galaxy is found to vary between $20-80$ pc in the inner disc, $R \lesssim 10$ kpc \citep{sanders84}, whereas the same flares up to $\sim 100-200$ pc at the outer radii \citep{grabelsky87}. In these regions, the molecular scale height closely matches the \HI~scale heights \citep[see, e.g., ][]{sodroski87,wouterloot90,kalberla08,kalberla09}. Nevertheless, our estimated scale heights for NGC 6946 is found to be consistent with several external galaxies. For example, \citet{scoville93} performed a CO aperture synthesis observation on the edge-on galaxy NGC 891 using the Owens Valley Millimeter Array with a spatial resolution of $\sim 2.3$\arcs~(106 pc). They estimated the molecular scale height in this galaxy to be $\sim 80$ pc at the center, which increases to $\sim 140$ pc at the outermost radius of their measurement, i.e., $\sim 10$ kpc. Their measurements are consistent with our estimates of thin molecular disc scale heights. Further, \citet{garcia-burillo92} used single-dish measurements with a larger beam of $\sim 13$\arcs~($\sim 600$ pc) in NGC 891 to find a larger molecular scale height of $\sim 200-300$ pc, which is consistent with the scale heights of our thick disc molecular gas.  Due to the diffuse nature of the thick disc gas, it is expected to be resolved out in the interferometric observation of \citet{scoville93}.

Recently, \citep{bacchini19} employed the hydrostatic equilibrium condition in a number of disc galaxies to determine the volume densities of different disc components along with star formation rates. Using these volume densities, they calculated the molecular and \HI~discs' scale height profiles in NGC 6946 (their Fig. 4). Their estimates of the scale heights in NGC 6946 is somewhat lower than what we find here. We note that to determine the volume densities, they did not solve the hydrostatic equilibrium equation in a self-consistent manner and ignored the gravity due to gaseous discs. Further, they considered the molecular disc to a single component system and assumed its velocity dispersion to be half that of the \HI. The \HI~velocity dispersions were determined by a 3D tilted ring fitting of the \HI~spectral cube and parametrized by an exponential function. For NGC 6946, thus computed molecular velocity dispersion values could be as low as 3 \kms~at radii $\gtrsim 5$ kpc. This is $\sim$ two times lower than what we have assumed for the thin disc (7 \kms). This implies that our scale height profiles at these radii are expected to be $\sim 2$ times larger than their values, which we observe here. 

Moreover, the \HI~scale height (same as the thick disc scale height) we find here for NGC 6946 is also consistent with other measurements in external galaxies. For example, early deep \HI~observations of NGC 891 reveals the existence of \HI~to significantly large heights from the midplane \citep[][see also, \citet{oosterloo07}]{swaters97}. \citet{zschaechner15b} used the Very Large Array (VLA) to observe the \HI~in the edge-on galaxy NGC 4013. By using a tilted ring model fitting to the observed \HI~spectral cube, they estimated an upper limit to the \HI~scale height to be $\sim 280$ pc at the central region, which increases to $\sim 1$ kpc at radii $\gtrsim 7$ kpc. Further, detailed kinematic modeling of \HI~in several edge-on galaxies using HALOGAS survey data allowed estimation of \HI~extent in the vertical direction \citep{kamphuis13,zschaechner12,zschaechner15}. These studies found the \HI~scale height in these galaxies to be a few hundred parsecs, which compares very well with what we estimated for NGC 6946.

The scale heights in NGC 6946 can vary depending on the assumed thick disc molecular gas fraction, $\rm f_{tk}$. The shaded and the hatched regions in Fig.~\ref{sclh} depict the variation of the scale heights for the molecular and \HI~discs for $\rm f_{tk}$ values between 0.3-0.9. As can be seen from the figure, for different $\rm f_{tk}$, the molecular scale height (thin or thick disc) changes by a few tens of parsecs. This, in turn, might not produce a significant difference in the total intensity (surface density) map, even for an edge-on orientation. Hence, an extremely high spatial resolution (tens of parsecs) would be required to distinguish molecular discs due to different $\rm f_{tk}$, which often is not available.

\subsection{Spectral cubes and width profiles}

Next, we use the density solutions, observed rotation curve, and the velocity dispersion profiles to construct a three-dimensional dynamical model of the molecular disc in NGC 6946. To do that, we construct a 3D grid with 1000 cells in each dimension. We chose a cell size of 25 pc, which is sufficient to eclose the molecular disc entirely ($\sim$ 19 kpc in diameter) and adequate to sample the molecular disc with sufficient density. We interpolate the molecular gas density (both thin and thick disc), rotation velocity, and velocity dispersions in this grid to create a complete 3D dynamical model. This grid is then inclined to the observed inclination (33$^o$, \citet{deblok08}), projected into the sky plane, convolved with the telescope beam (13.4\arcs $\times$ 13.4\arcs, \citet{leroy09a}), and sampled to the observed velocity resolution of 2.6 \kms~\citep{leroy09a} to produce an observation equivalent CO spectral cube. We repeat this exercise for molecular discs with different assumed $\rm f_{tk}$ and consecutively produce several CO spectral cubes. These spectral cubes are then compared with the observed one to investigate the properties of the thin and the thick molecular discs in NGC 6946.

\begin{figure}
\begin{center}
\begin{tabular}{c}
\resizebox{0.45\textwidth}{!}{\includegraphics{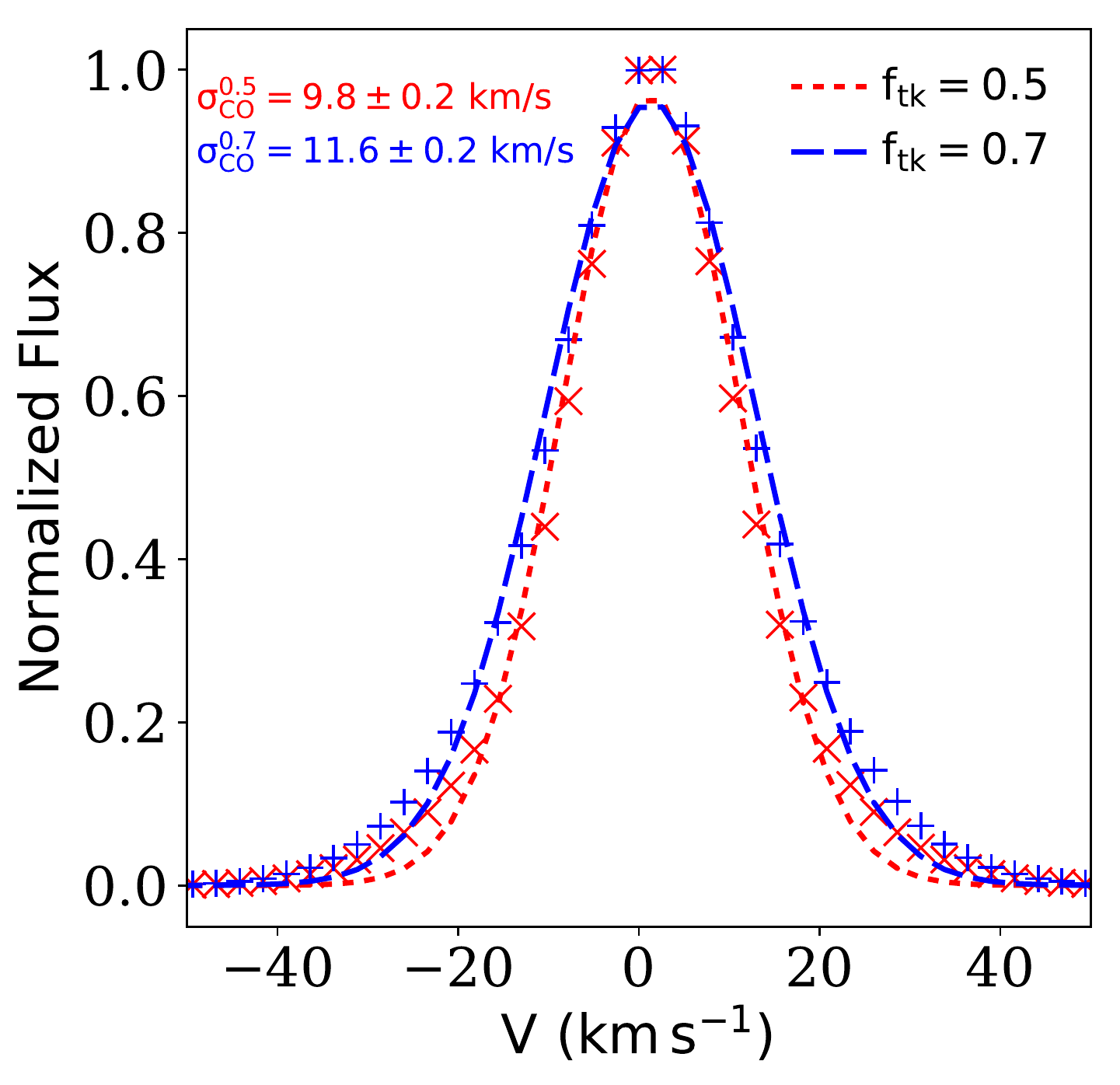}}
\end{tabular}
\end{center}
\caption{Simulated normalized stacked CO spectra for NGC 6946 at a radius of $\sim$ 6 kpc. The red crosses represent the stacked spectrum for a $\rm f_{tk} = 0.5$, whereas the blue pluses denote the same for a $\rm f_{tk} = 0.7$. Single-Gaussian fits to the spectra are also shown by the red and the blue dashed lines, respectively. The corresponding $\sigma_{CO}$ values for thick disc molecular gas fraction of 0.5 and 0.7 are found to be 9.8 \kms~and 11.6 \kms, respectively.}
\label{stack}
\end{figure}

\begin{figure*}
\begin{center}
\begin{tabular}{c}
\resizebox{1.\textwidth}{!}{\includegraphics{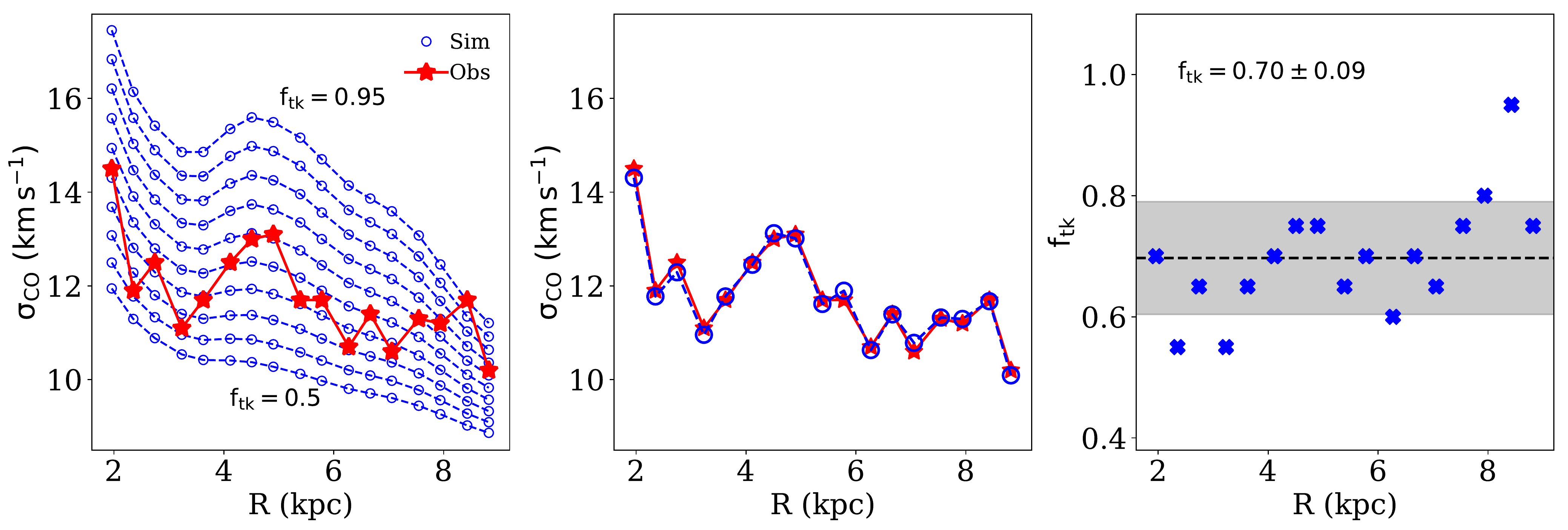}}
\end{tabular}
\end{center}
\caption{The left panel shows the comparison of the simulated $\sigma_{CO}$ profiles with the observed one for NGC 6946. The solid red stars represent the observed $\sigma_{CO}$ as obtained by stacking the CO spectra by \citet{calduprimo13}. The empty blue circles represent the simulated $\sigma_{CO}$ profiles for different assumed thick disc molecular gas fraction. The simulated $\sigma_{CO}$ profile at the top is for a $\rm f_{tk} = 0.95$ and the bottom one is for $\rm f_{tk}=0.5$. In between, the profiles are for a $\rm f_{tk}$ values separated by 0.05. The middle panel shows the best fit simulated $\sigma_{CO}$ profile. The solid red stars represent the observed $\sigma_{CO}$ profile, whereas the empty blue circles show the best fit simulated profile. Right panel: Shows the thick disc molecular gas fraction at every radial bin, which produces a simulated $\sigma_{CO}$ profile that matches the observation best. The black dashed line represents the mean $\rm f_{tk} = 0.70$, whereas the gray shaded region represents the 1-$\sigma$ error on this mean value (0.09). As can be seen, our estimated $\rm f_{tk}$ values show a moderate incremental trend as a function of radius (with a Spearman rank correlation coefficient of 0.6, see the text for more details).}
\label{thick_frac}
\end{figure*}

Using the same definition as used by \citet{calduprimo13}, we define the spectral width of a spectrum as the width (sigma) of a single-Gaussian function fitted to it. We use the same strategy to estimate the spectral widths in the simulated CO discs, as adopted by \citet{calduprimo13}, \citep[see also,][]{patra18c,patra20a}. We first identify all the synthetic spectra within a radial bin of width equal to the observing beam. These spectra are then fitted with Gaussian-Hermite Polynomials of order three to locate their centroids. All the centroids of the spectra are then shifted to a common velocity. This, in turn, aligns all the spectra, after which we stack them to produce a stacked spectrum. Thus we produce several stacked spectra for all the radial bins. These stacked spectra are then fitted with single-Gaussian functions to estimate the spectral widths. Using the observed CO spectral cube (from the HERACLES survey), we estimate the stacked SNR in the same radial bins of our synthetic cube. Subsequently, we add an equivalent amount of noise to the simulated stacked spectra. These stacked spectra now can be considered to be observation equivalent. We bootstrap the stacked spectrum at every radial bin by producing its 1000 realizations with the same noise properties (but different noise values). All these 1000 spectra are then fitted with single-Gaussian functions. The standard deviation of their widths is then added to the fitting error in quadrature to estimate the error on the spectral width at a radial bin. Thus we produce a spectral width profile from a simulated spectral cube. Several such spectral width profiles are generated for molecular discs with different assumed $\rm f_{tk}$. In Fig.~\ref{stack}, we show two representative stacked spectra for different $\rm f_{tk}$ of 0.5 (red crosses) and 0.7 (blue pluses) at a radius of $\sim$ 6 kpc. For better comparison, we normalize both the spectra to unity. Single-Gaussian fits to the spectra are shown by the broken lines in the figure. As can be seen, different $\rm f_{tk}$ values result in stacked spectra with different spectral widths. We find a $\sigma_{CO} = 9.85 \pm 0.18$ \kms~for $\rm f_{tk} = 0.5$ and a $\sigma_{CO} = 11.65 \pm 0.19$ \kms~for $\rm f_{tk} = 0.7$ which are readily distinguishable. As can also be seen from the figure, single Gaussian fits to the stacked spectra fail to capture the broad wings produced by the thick disc molecular gas. In that sense, a single Gaussian fit would be less sensitive to the presence of thick disc molecular gas in a spectrum. To examine the same, we fitted double Gaussian profiles to the stacked spectra and used the broad component's width as an indicator of the thick disc molecular gas fraction. However, we find that a double Gaussian fitting deblends the thin and thick disc's spectral widths. As a result, the width of the broad component (or the narrow component) renders the CO velocity dispersion insensitive to $\rm f_{tk}$ and so isn't used here. In Fig.~\ref{thick_frac} left panel, we plot the simulated $\sigma_{CO}$ profiles (empty blue circles connected by blue dashed lines) for $\rm f_{tk} = 0.5$ to 0.95 in steps of 0.05 (though we run our simulations for $\rm f_{tk} = 0.05-0.95$). For comparison, we also show the observed $\sigma_{CO}$ profile (solid red stars connected by an unbroken red line) as estimated by \citet{calduprimo13}. As can be seen from the panel, our choice of $\rm f_{tk}$ values produces simulated $\sigma_{CO}$ such as it encloses the observed $\sigma_{CO}$ completely.

\subsection{CO velocity dispersion and thick disc fractions}

Next, using the simulated $\sigma_{CO}$ profiles, we perform a $\chi^2$ optimization at every radial point (defined by the radius of the observed $\sigma_{CO}$) and search for the thick disc molecular gas fraction, $\rm f_{tk}$ which produces a $\sigma_{CO}$ closest to the observed value at that radius. Thus we obtain the $\rm f_{tk}$ values as a function of radius, which produces a $\sigma_{CO}$ profile best matched to the observed one. It should be mentioned here that, as the spectral widths are evaluated in bins a beam apart, and the spectral blending in NGC 6946 is minimal, there is no considerable correlation between the spectral widths in consecutive radial bins. Hence, the best fit $\rm f_{tk}$ value can be retrieved in every radial bins separately. In the middle panel of Fig.~\ref{thick_frac}, we show thus obtained the best simulated $\sigma_{CO}$ profile (empty blue circles connected with a blue dashed line) along with the observed one (solid red stars connected by an unbroken red line). As can be seen from the panel, the simulated and the observed $\sigma_{CO}$ profiles match closely, indicating satisfactory modeling of the molecular discs in NGC 6946. The corresponding best fit $\rm f_{tk}$ profile is plotted in the right panel of Fig.~\ref{thick_frac}. The optimized $\rm f_{tk}$ was found to vary between 0.6-0.95 with a mean value of 0.70$\pm$0.09. As can be seen from the panel, the $\rm f_{tk}$ values in NGC 6946 across its molecular disc appears to be consistent with the mean value with a minimal variation. To test its dependence on the radius quantitatively, we evaluate the Spearman rank correlation coefficient. We find the correlation coefficient to be 0.6. This indicates a moderate increment of $\rm f_{tk}$ as a function of radius.  

\begin{figure*}
\begin{center}
\begin{tabular}{c}
\resizebox{1.01\textwidth}{!}{\includegraphics{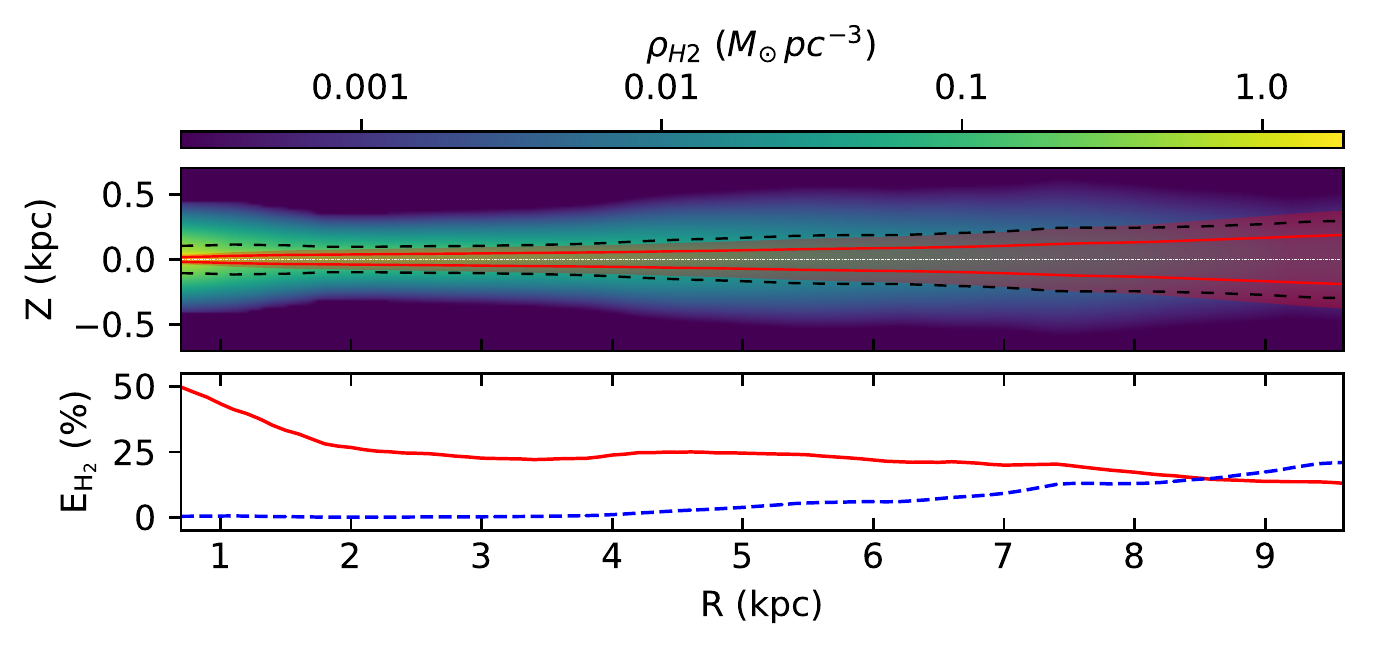}}
\end{tabular}
\end{center}
\caption{Top panel: The color scale shows the distribution of the molecular gas in the vertical direction as a function of radius. The white dotted line represents the midplane. The solid red lines and the dashed black lines represent the thin and thick disc scale heights, respectively. The red shaded region encloses twice the scale height of the thin disc. Any molecular gas outside this region is considered extra-planar. Bottom panel: The fractional amount of extra-planar molecular gas as a function of radius. The solid red line represents the extra-planar molecular gas fraction adopting twice the thin disc scale height, as shown in the top panel, whereas the blue dashed line indicates the fraction for a constant scale height of 300 pc.}
\label{egas}
\end{figure*}

\begin{figure}
\begin{center}
\begin{tabular}{c}
\resizebox{0.45\textwidth}{!}{\includegraphics{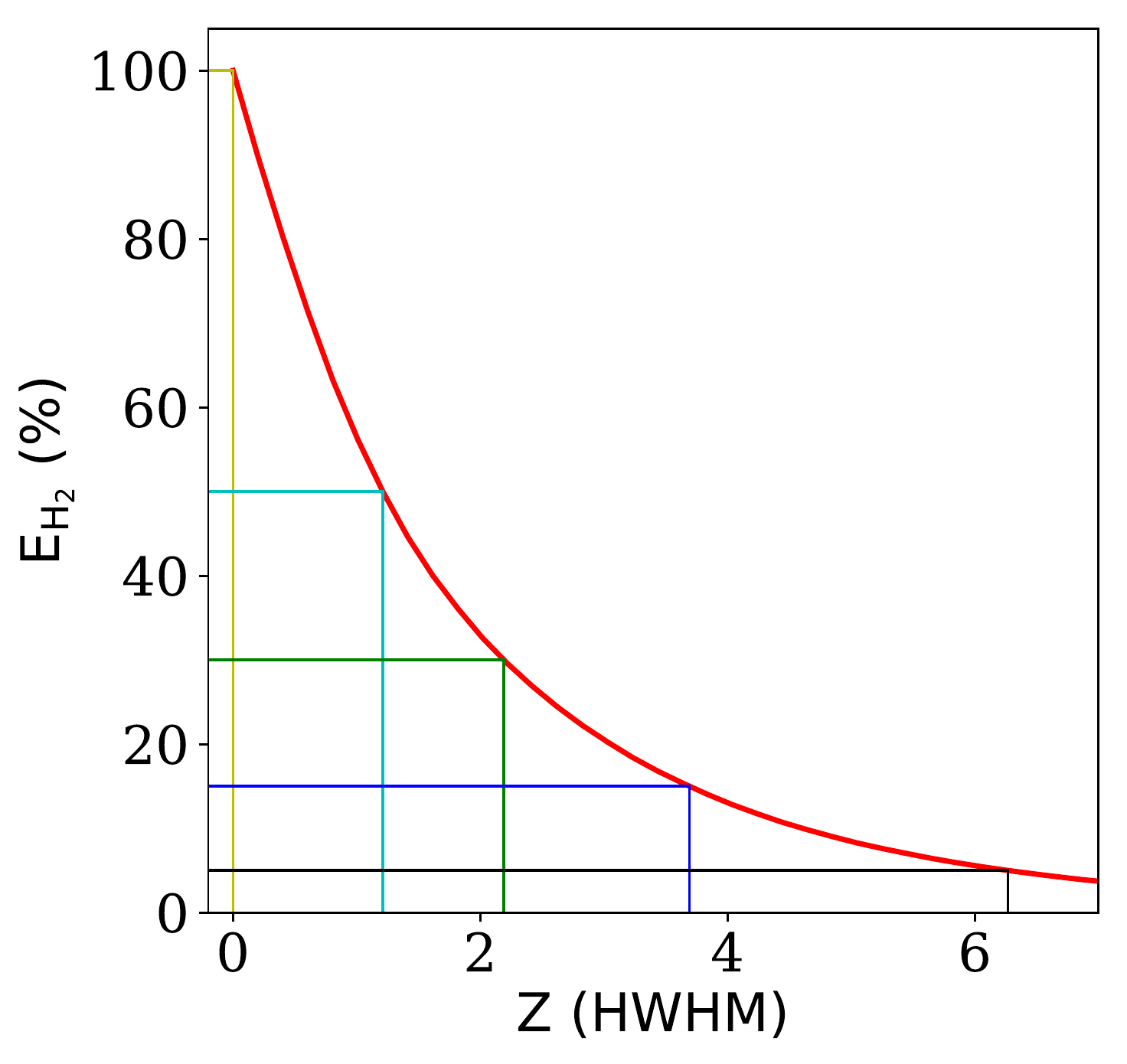}}
\end{tabular}
\end{center}
\caption{Fractional molecular mass in NGC 6946 above an altitude. The solid red line represents the fraction of the total molecular mass (for the entire galaxy) located above a given altitude, $\rm Z$ measured in the units of the thin disc scale height. The yellow, cyan, green, blue, and black lines demarcate respectively altitudes where 100\%, 50\%, 30\%, 15\%, and 5\% molecular gas are located. In NGC 6946, $\sim$ 32\% molecular gas is located above twice the thin disc scale height. This gas is considered to be extra-planar.}
\label{egas_frac}
\end{figure}

This result is consistent with the finding of \citet{pety13}, where they concluded at least 50\% of the molecular gas in M51 is in the thick disc. Not only the thick disc molecular gas fraction, but the molecular scale height we find here (20-300 pc, Fig.~\ref{sclh}) for NGC 6946 is also consistent with what was found by \citet{pety13} for M51. In an earlier study, however, \citet{garcia-burillo92} observed the edge-on galaxy NGC 891 in CO and found the extent of its molecular disc to be up to 1-1.4 kpc from the midplane. A Gaussian fitting to the observed intensity profile in the vertical direction results in a vertical scale height of $\sim 600$ pc, which is $\sim 2$ times larger than what we find here. \citet{garcia-burillo92} estimated the amount of molecular gas in the thick disc to be $\sim$ 20\% of the total gas content in NGC 891. However, they also point out that $\sim$ 40\%-60\% gas in NGC 891 should be in the form of low density, high velocity dispersion to explain the typical ratio of the $\rm ^{12}CO \ (1 \rightarrow 0)$ to $\rm ^{13}CO \ (1 \rightarrow 0)$ brightness. Later, \citet{sofue93} imaged NGC 891 in more detail and found CO emissions at heights of $\sim$ 3.5 kpc from the center. They concluded that the amount of thick disc molecular gas in NGC 891 could be as high as 50\%. \citet{roman-duval16} used $^{12}CO$ and $^{13}CO$ observations of the Galaxy to find $\sim 10-20\%$ of the total molecular gas in the diffused state inside a radius of $\sim 3-4$ kpc, whereas this fraction increases to $\sim 50\%$ at $R \sim 15$ kpc. Further, in a recent study, \citet{fumiya20} used $^{12}CO$ observations of the bar region in the galaxy NGC 1300 to find $\sim 75-90\%$ of the molecular gas in the diffuse thick disc. This fraction reduces to $\sim 30-65\%$ in the arm and bar-end region. Our result of $\sim 70 \pm 09\%$ thick disc molecular gas in NGC 6946 thus largely consistent with these previous studies.

\subsection{Origin of extra-planar molecular gas}

The estimation of the thick disc molecular gas fraction has many implications. For example, one can estimate the amount of extra-planar molecular gas in NGC 6946. The extra-planar molecular gas ($\rm E_{H_2}$) is defined as the gas, which lies above twice the scale height of the thin disc \citep[see, e.g.,][]{marasco11}. We use the optimized density solutions (taking care of the appropriate optimized $\rm f_{tk}$ at every radius) to estimate the thin disc molecular scale height at every radius and calculate the fractional amount of the molecular gas above twice this scale height. In Fig.~\ref{egas} top panel, we show the distribution of the molecular gas (thin+thick) in the $R-z$ plane. The thin (solid red line) and thick (black dashed line) disc scale heights are also shown. The red shaded region demarcates twice the thin disc scale height. Any molecular gas outside this region is considered extra-planar. It can be seen from the figure that the amount of $\rm E_{H_2}$ at any radius strongly depends on the molecular scale height (of the thin disc) at that radius. The bottom panel of the figure shows the $\rm E_{H_2}$ fraction as a function of radius. At inner radii, much lower scale heights result in a significantly high $\rm E_{H_2}$ fraction. Whereas, at outer radii, the $\rm E_{H_2}$ fraction steadily decreases with increasing scale height. For NGC 6946, $\rm E_{H_2}$ fraction vary from $\sim 50\%$ at the inner radii to $\sim 15\%$ at the outer radii with a mean of $23.3 \pm 7.5 \ \%$. For comparison, we also estimate the $\rm E_{H_2}$ fraction (blue dashed line in the bottom panel of Fig.~\ref{egas}) for a constant scale height of 300 pc (scale height of the thin disc molecular gas in NGC 891, \citet{sofue93}). As can be seen, the $\rm E_{H_2}$ fraction for this case increases as a function radius from less than a few percent at the inner radii to $\sim 20\%$ at the edge of the molecular disc. This indicates that the flaring of the molecular disc should be taken into account while estimating the $\rm E_{H_2}$ fraction. Using the thin disc scale height profile to define $\rm E_{H_2}$, we find that the thick disc dominantly contributes to the extra-planar molecular gas. The thin disc only contributes $\sim 5\%$ at the inner disc, which increases to $\sim 20\%$ at the edge.

To further investigate how the total molecular mass in NGC 6946 is distributed above the plane, in Fig.~\ref{egas_frac}, we plot the total molecular gas fraction above an altitude (defined in the units of the thin disc scale height). This represents the cumulative fraction of the molecular gas considering contributions from all radii. In NGC 6946, we find that $\sim 32\%$ of the total molecular gas is located above twice the thin disc's scale height. This fraction steadily decreases and falls below 10\% above an altitude of $\sim 4$ times the thin disc scale height. This extra-planar molecular gas fraction is consistent with what is found by \citet{pety13}. They found that $\sim 33\%$ of the total molecular gas in M51 is located above an altitude twice the thin disc (dense gas in their description) scale height (see their Fig. 20). In a recent study, \citet{marasco19} used deep interferometric \HI~data from the HALOGAS survey to investigate the extra-planar neutral gas in 15 nearby galaxies. They found that the extra-planar \HI~is nearly ubiquitous and contributes $\sim 5-25\%$ to the total neutral gas in galaxies. Albeit for neutral gas, their estimated extra-planar gas fraction compares well with our results.

The origin of this thick extra-planar molecular gas, however, is not well understood. Several stellar processes in the discs of galaxies could provide possible mechanisms, e.g., galactic fountains or chimneys, to produce extra-planar gas \citep{shapiro76,bregman80,putman12}. In fact, several high-resolution studies of nearby galaxies found direct evidence of starburst-driven expulsion of molecular gas to significant heights. For example, using high resolution ($\sim$ 70 pc) observation of $\rm ^{12}CO \ (1 \rightarrow 0)$, \citet{walter02} found molecular streamers in M82. These streamers are believed to be closely related to the starburst of the galaxy and can expel a significant amount of molecular gas to a height $\sim $ 1.2 kpc from the midplane. Likewise, several other studies have also found ample evidence of starburst-driven molecular outflows reaching about a few kpc above the midplane \citep{weiss99,sakamoto06,feruglio10,bolatto13}. Further, AGN activity in the central regions of galaxies could also play a vital role in ejecting a significant amount of molecular gas into high altitudes \citep{alatalo11,sturm11}. Though NGC 6946 does not host any AGN \citep{tsai06}, it shows significantly high star formation activity in its optical disc \citep{degioia84} with signs of many \HII~complexes even beyond the optical ($r_{25}$) radius \citep{ferguson98}. This enhanced star formation activity could possibly explain the estimated large amount of extra-planar molecular gas. In addition to these internal sources, accretion of gas from the CGM can also be a viable mechanism to produce a substantial amount of extra-planar molecular gas \citep{oort70,binney05,fraternali05,kaufmann06}. In fact, \citet{boomsma08} studied the extra-planar \HI~in NGC 6946 and concluded that while stellar feedback mechanisms could explain the extra-planar \HI~at the inner radii, gas accretion could play a vital role in producing the same at outer radii. However, to further understand the origin and sustenance of this extra-planar/thick disc molecular gas in galaxies, one needs to perform a similar analysis as presented here to a larger sample of galaxies, which we plan to do next.

\section{conclusion}

In conclusion, we model the molecular disc in NGC 6946 as a two-component system with a thin disc having a low velocity dispersion of 7 \kms~and a thick disc with a velocity dispersion equal to that of the \HI. With this, we model the baryonic disc in NGC 6946 as a four-component system consisting of a stellar disc, \HI~disc, and two molecular discs (thin and thick). We assume that these discs are in vertical hydrostatic equilibrium under their mutual gravity in the external force-field of the dark matter halo. Under this assumption, we formulate the Poisson-Boltzmann equation of hydrostatic equilibrium and solve it numerically using an 8$^{th}$ order Runge-Kutta method to obtain a three-dimensional density distribution of the baryonic discs. In particular, we investigate the three-dimensional density distribution of the molecular gas in NGC 6946 and its implications on the observed CO spectral widths. 


Inspecting the solutions of the hydrostatic equilibrium equation, we find that the density of the individual disc components deviates from a conventional $sech^2$  law due to the coupling between the disc components. In the presence of other disc components, the density solutions behave more like a Gaussian function. We also find that for NGC 6946, the \HI~and the thick disc molecular gas acquires a larger scale height as compared to the thin disc molecular gas. The molecular scale height in the thin disc was found to vary between $20-200$ pc, whereas the same found to vary between $100-300$ pc in the thick disc. Due to the same assumed velocity dispersions, the thick disc molecular gas and the \HI~shows the same scale height. 

We further use the density solutions, the observed rotation curve, and the velocity dispersion profiles to build three-dimensional dynamical models of molecular discs in NGC 6946. These dynamical models then inclined to the observed inclination, projected into the sky plane, and convolved with the telescope beam to produce observation equivalent simulated CO spectral cubes. A number of such spectral cubes are produced for different assumed $\rm f_{tk}$ in the molecular disc of NGC 6946. These spectral cubes are further used to produce CO spectral width profiles by stacking CO spectra within radial bins and fitting them with single-Gaussian functions. These spectral width profiles are then compared with the observed one to constrain the thick disc molecular gas fraction in NGC 6946.

Using a $\chi^2$ method, we estimate the thick disc molecular gas fraction in NGC 6946 at all the radial bins, which produces a simulated spectral width that best matches the observation. We find that the best optimized simulated spectral width profile matches the observed one very well. We also find that the corresponding thick disc molecular gas fraction profile almost remains constant (albeit with a slight sign of increasing trend) as a function of the radius with a mean value of $0.70 \pm 0.09$. This means that in NGC 6946, $\sim$ 70\% of the total molecular gas is in the thick disc. \emph{This is the first-ever estimation of the thick disc molecular gas fraction in an external galaxy as a function of the radius with sub-kpc spatial resolution.}

We also estimate the amount of extra-planar molecular gas in NGC 6946. The extra-planar molecular gas ($\rm E_{H_2}$) is defined as the gas, which lies above twice the scale height of the thin disc \citep[see, e.g.,][]{marasco11}. The fraction of extra-planar molecular gas in NGC 6946 was found to vary from $\sim 50\%$ at the center to $\sim 15\%$ at the edge of the molecular disc with a mean value of $\sim 23.3\% \pm 7.5\%$. A high fraction of the extra-planar molecular gas at the center of NGC 6946 supports the idea that this extra-planar gas could originate through different stellar feedback processes, e.g., galactic fountain or chimney. However, understanding the origin and sustenance of the thick disc/extra-planar molecular gas in galaxies would require a similar study in a larger sample of galaxies, which we plan to do next.

\section{Acknowledgement}
This work made use of The \HI~Nearby Galaxy Survey (THINGS, \citet{walter08}) and The HERA CO-Line Extragalactic Survey (HERACLES, \citet{leroy09a}) data. NNP would like to thank the THINGS and the HERACLES team for making the data publicly available. 

\section{Data availability}
We used already existing publicly available data for this work. No new data were generated or analyzed in support of this research. 

\bibliographystyle{mn2e}
\bibliography{bibliography}

\end{document}